# New anti-perovskite-type Superconductor $ZnN_yNi_3$

Masatomo UEHARA[*], Akira UEHARA, Katsuya KOZAWA, and Yoshihide KIMISHIMA.

*Department of Physics, Faculty of Engineering, Yokohama National University, Hodogaya-ku Yokohama 240-8501*

**Abstract**

We have synthesized a new superconductor $ZnN_yNi_3$ with $T_c$ ~3 K. The crystal structure has the same anti-perovskite-type such as $MgCNi_3$ and $CdCNi_3$. As far as we know, this is the third superconducting material in Ni-based anti-perovskite series. For this material, superconducting parameters, lower critical field $H_{c1}(0)$, upper critical field $H_{c2}(0)$, coherence length $\xi(0)$, penetration depth $\lambda(0)$, and Gintzburg-Landau parameter $\kappa(0)$ have been experimentally determined.

## 1. Introduction

Since He *et al.* have found a new superconductor MgCNi$_3$, having the $T_c$ of ~ 8 K, much attention has been paid to anti-perovskite superconductor. This material includes a lot of ferromagnetic Ni and has a structural similarity with fcc elemental Ni.[1], implying that the ferromagnetic correlation is associated with the superconductivity of MgCNi$_3$. Several experiments to reveal the superconducting-gap symmetry have been carried out[2-5], but the consensus on gap symmetry and the origin of superconductivity has not been obtained so far. A theoretical calculation has pointed out that this compound is located near the ferromagnetism due to the existence of a large peak in the density of states (DOS) slightly below Fermi energy ($E_F$) and predicted the possibility of inducing ferromagnetism by hole doping[6]. Hole doping causes an increase in DOS at $E_F$ ($D(E_F)$) assuming the rigid-band picture. Actually the emergence of ferromagnetism has been reported in carbon-deficient (Mg,Zn)C$_y$Ni$_3$ ($y$<0.7)[7]. Several analogous materials to MgCNi$_3$ have been synthesized but only CdCNi$_3$ showed superconductivity[8]. Many Ni-based anti-perovskite materials are ferromagnets for example (Mg,Zn)C$_y$Ni$_3$ ($T_C$=150~300 K for $y$ < 0.7)[7], AlC$_y$Ni$_3$ ($T_C$=41.5 and 300 K for $y$=0 and 1)[9,10], ZnC$_y$Ni$_3$ ($T_C$=180 and 30 K for $y$=0 and 0.5)[7] and In$_{0.95}$CNi$_3$ ($T_C$=577 K)[11]. GaC$_y$Ni$_3$ ($y$=0 and 1)[9,12] shows the exchange-enhanced Pauli paramagnetism[13], indicating the existence of strong ferromagnetic correlation. These facts mentioned above indicate that Ni-based anti-perovskite materials tend to have ferromagnetic character. In order to clarify the relationship between superconductivity and ferromagnetism in Ni-based anti-perovskite superconductor, it has been hoped to find new superconductor in addition to MgCNi$_3$ and CdCNi$_3$.

In this paper, we report the synthesis and superconducting parameters of newly synthesized ZnN$_y$Ni$_3$ with $T_c$ ~3 K, lower-critical field $H_{c1}(0)$, upper-critical field $H_{c2}(0)$, coherence length $\xi(0)$, penetration depth $\lambda(0)$, and Gintzburg-Landau parameter $\kappa(0)$.

## 2. Experimental

A series of samples was prepared with elemental Zn and Ni powders. The powders were weighed with a nominal composition of $Zn_{1.05}Ni_3$ and mixed, then pressed into pellet. The extra Zn powder was added to compensate vaporizing quantities. The pellet was sintered in $NH_3$ gas at 400 ˚C for 3 h, 520 ˚C for 15h, and 600 ˚C for 5 h with intermediate grindings. $NH_3$ gas decomposes to chemically active hydrogen and nitrogen at high temperature and active nitrogen penetrates into the sample and the sample is nitrified. This has been known as an effective method to make 3$d$-transition metal nitride[14-16].

X-ray diffraction patterns were obtained by using CuK$_\alpha$ radiation. Electrical resistivity was measured by standard four-probe method and the magnetization measurements were performed by using a Quantum Design SQUID magnetometer. Magnetization was measured with a field of 10 Oe after zero-field cooling (ZFC) and field cooling (FC). No demagnetic correction was applied.

## 3. Result and Discussion

Fig. 1 shows the powder X-ray diffraction patterns of $ZnN_yNi_3$. All diffraction patterns can be indexed by a cubic structure with space group of *Pm3m*. The lattice parameter and nitrogen content $y$ are refined by Rietveld analysis with Rietan program. The lattice parameter and nitrogen content are determined to be 3.756 Å and 1.012± 0.208. Large deviation of nitrogen content is attributed to the small scattering cross-section of nitrogen, inherent problem of conventional X-ray source. The lattice parameter of $ZnN_yNi_3$ is close to those predicted by calculations for stoichometric $ZnCNi_3$ (3.74[17] and 3.72 Å[18]). Considering that the covalent radius of nitrogen and carbon is almost identical, $ZnN_yNi_3$ is likely to be stoichometric, i.e., $y$~1. It should be noted that the lattice parameter of $ZnCNi_3$ has been reported experimentally as 3.66 Å and this small value is probably due to carbon-defect[17,19]. The nitrogen content $y$ can also be estimated from the sample weight change between before and after sintering. The weight change is ~104 %, corresponding to $y$ ~ 0.98. These two estimated values of $y$=1.012 and 0.98 are roughly consistent, so that $y$ is likely to locate around 1. In order to obtain accurate $y$ value, higher quality diffraction data such as syncrotron data, or some sort of chemical analysis are needed. Rietveld refinement intensities are denoted with + marks and the difference between the data and calculation is shown at the bottom in Fig. 1. The expected Bragg peak positions are indicated by the vertical lines in Fig. 1.

By resistivity and magnetization measurements shown in Fig. 2(a)(b), the sample has been revealed to be a superconductor with $T_c$ (onset) ~ 3.0 K. The metallic conduction is seen from 3 K up to 300 K (see inset of Fig. 2(a)). In the magnetization measurement (Fig. 2(b)), clear shielding effect by Meissner diamagnetism is observed in ZFC data. The superconducting volume fraction estimated from ZFC data at 1.8 K is 186 %, showing the bulk nature of this superconductivity. The very large volume fraction value of 186 % may be due to demagnetization effect. Comparing with ZFC data, the diamagnetic signal in FC data is small, indicating that the sample is type-II superconductor with flux pinning centers in it.

In order to determine lower critical field $H_{c1}(0)$ and upper critical field $H_{c2}(0)$, magnetization measurements as a function of applied field at several fixed temperatures below $T_c$ and electrical resistivity measurements under various magnetic fields have been carried out. The inset in Fig. 3(a) shows the applied field dependence of magnetization. The magnetic field at which $M$-$H$ curves deviates from a linear relation is defined to be $H_{c1}(T)$ at a given temperature as shown by arrows in the inset of Fig. 3(a). In Fig. 3(a), $H_{c1}(T)$ is plotted as a function of temperature. The data of $H_{c1}(T)$ are fitted with the empirical formula $H_{c1}(T)=H_{c1}(0) \cdot (1-(T/T_c)^2)$ and $H_{c1}(0)$ has been determined to be 6.9 mT.

The inset of Fig. 3(b) shows the resistivity data as a function of temperature under various magnetic fields. $T_c$ has been defined as the maximum of $d\rho/dT$. As shown in this figure, $T_c$ decreases with increasing magnetic field and the superconducting transition almost disappears below 0.8 T. The magnetic field providing a certain $T_c(H)$ is defined as $H_{c2}(T)$ ($T = T_c$). In Fig. 3(b), the temperature dependence of $H_{c2}(T)$ is plotted. On the basis of this plot, $H_{c2}(0)$ has been calculated using the Werthamer-Helfand-Hohenberg (WHH) formula[20],

$$H_{c2}(0) = -0.693 \cdot (dH_{c2}/dT)_{T=T_c} \cdot T_c.$$

The determined $H_{c2}(0)$ is 0.96 T, using the WHH formula (see Fig. 3(b)). The $H_{c2}(0)$ value actually corresponds to the magnetic fields at which superconductivity almost disappears (see insets of Fig. 3(b)). Using $T_c$=3 K, the Pauli limit value is calculated to be 5.5 T. The $H_{c2}(0)$ value (0.96 T) is suppressed from Pauli limit much more than in the case of MgCNi$_3$ ($H_{c2}(0)$=14.4 T and Pauli limit=14 T[4]). This is the case for CdCNi$_3$ ($H_{c2}(0)$=2.2 T and Pauli limit=6.2 T[18]). One of the possible reason for small $H_{c2}(0)$ observed in ZnN$_y$Ni$_3$ and CdCNi$_3$ is that the ferromagnetic correlation develops and causes spontaneous pair breaking. The superconducting coherence length $\xi(0)$ and penetration depth $\lambda(0)$ have been estimated by using the relation, $\xi(0) = (\phi_0 / 2\pi H_{c2}(0))^{1/2}$ and $\lambda(0) = (\phi_0 / \pi H_{c1}(0))^{1/2}$, where $\phi_0$ is the fluxoid. $\xi(0)$ and $\lambda(0)$ is 185 and 3089 Å, respectively. The Ginzberg-Landau parameters $\kappa = \lambda(0) / \xi(0)$ is 16.7, showing that this system is a type-II superconductor. Table I shows all the estimated superconducting parameters of ZnN$_y$Ni$_3$ and, for comparison, those of MgCNi$_3$ and CdCNi$_3$ from ref. 4 and 18 are also shown.

Analogous family compound ZnCNi$_3$ is not a superconductor even though the band calculation predicts the similar band structure with MgCNi$_3$[17]. The origin of discrepancy is considered to be a sort of "under-doped" state due to the existence of carbon-defect. In contrast, ZnN$_y$Ni$_3$ is thought to be "over-doped" with electron because nitrogen probably provides one more electron than carbon. Assuming the similar band structure and rigid-band picture, D($E_F$) of ZnN$_y$Ni$_3$ should be suppressed rather than those of MgCNi$_3$ and CdCNi$_3$. Therefore, $T_c$ might increase by reducing nitrogen content, or $T_c$ is already tuned to be optimum by nitrogen-defect. Precise chemical analysis of nitrogen is necessary to address mentioned above.

Finally, we would like to point out an issue as follows: ZnNi$_3$, ZnC$_{0.5}$Ni$_3$ and ZnN$_{0.5}$Co$_3$ are ferromagnets and it is easy to synthesize these samples with good quality[7,21]. Therefore, it can be expected to synthesis Zn(C,N)$_y$(Ni,Co)$_3$ solid solution. By studying carefully transform from ferromagnetic to superconducting phase with using Zn(C,N)$_y$(Ni,Co)$_3$, the crucial information about the relationship between ferromagnetism and superconductivity might be obtained in Ni-based anti-perovskite system.

In summary, a new superconductor ZnN$_y$Ni$_3$, which has the same anti-perovskite structure as MgCNi$_3$ and CdCNi$_3$, has been synthesized. The $T_c$ is ~3.0 K. Rietveld analysis and the weight change between before and after sintering indicate that the nitrogen content $y$ is nearly 1. By electrical resistivity and magnetization measurements, the superconducting parameters $H_{c1}(0)$, $H_{c2}(0)$, $\xi(0)$, $\lambda(0)$, and $\kappa(0)$ have been determined, revealing that ZnN$_y$Ni$_3$ is a type-II superconductor.

**Acknowledgements**

This work was partly supported by a Grant-in-Aid for Scientific Research from The Ministry of Education, Culture, Sports, Science and Technology, Japan and by Research Institute of Yokohama National University.

**Figure captions**

Figure 1: Powder X-ray diffraction patterns for $ZnN_yNi_3$. Rietveld refinement result is denoted with + marks and the difference between the data and calculation is shown at the bottom. The expected Bragg peak positions are indicated by the vertical lines.

Figure 2 (a): Temperature dependence of electrical resistivity $\rho$ for $ZnN_yNi_3$. The inset shows the temperature dependence $\rho$ from 1.8 up to 300 K.

Figure 2 (b): Temperature dependence of $\chi$ at 10 Oe between 1.8 and 5 K with FC and ZFC methods.

Figure 3 (a): Temperature dependence of $H_{c1}$. The inset shows *M-H* curves at several temperature used for determining $H_{c1}$ for $ZnN_yNi_3$.

Figure 3 (b): The relationship of $T_c$ vs $H_{c2}(T=T_c)$. The inset shows the temperature dependence of $\rho$ under several magnetic fields for $ZnN_yNi_3$. The magnetic field providing a certain $T_c(H)$ is defined as $H_{c2}(T)$ ($T = T_c$).

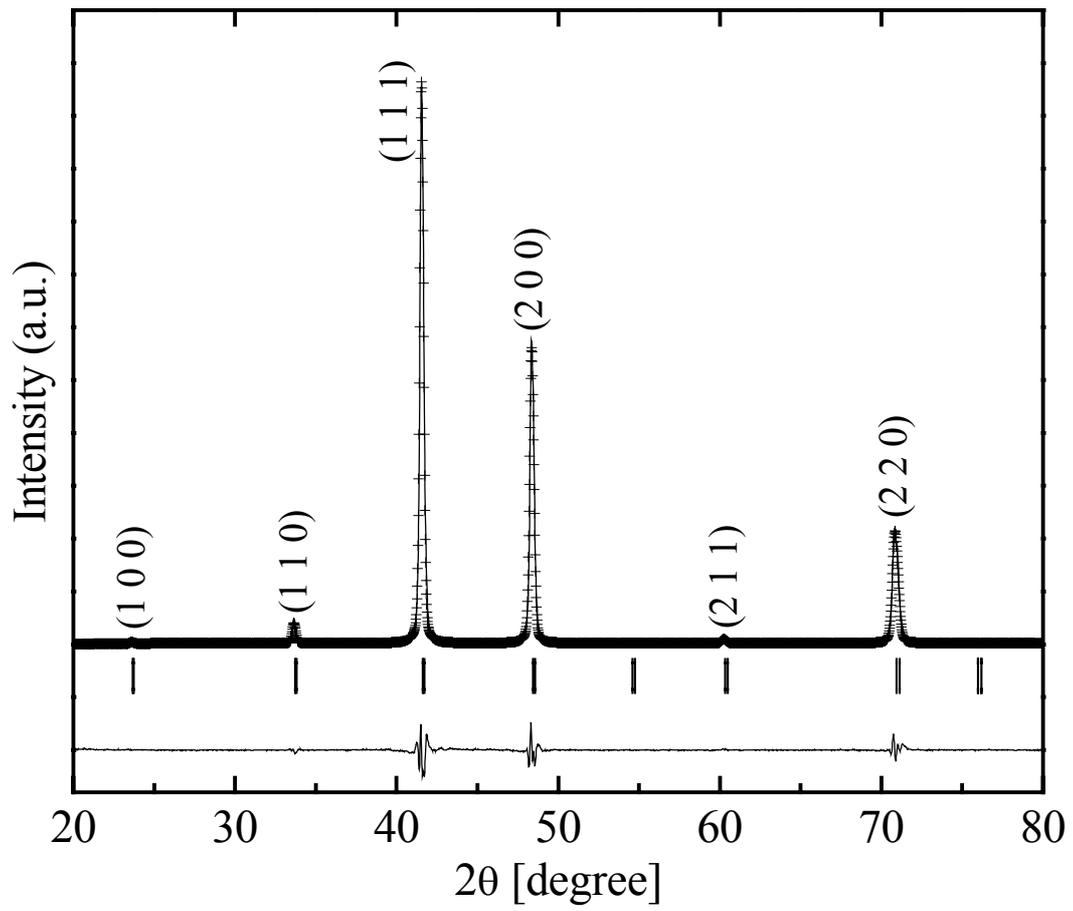

Fig. 1

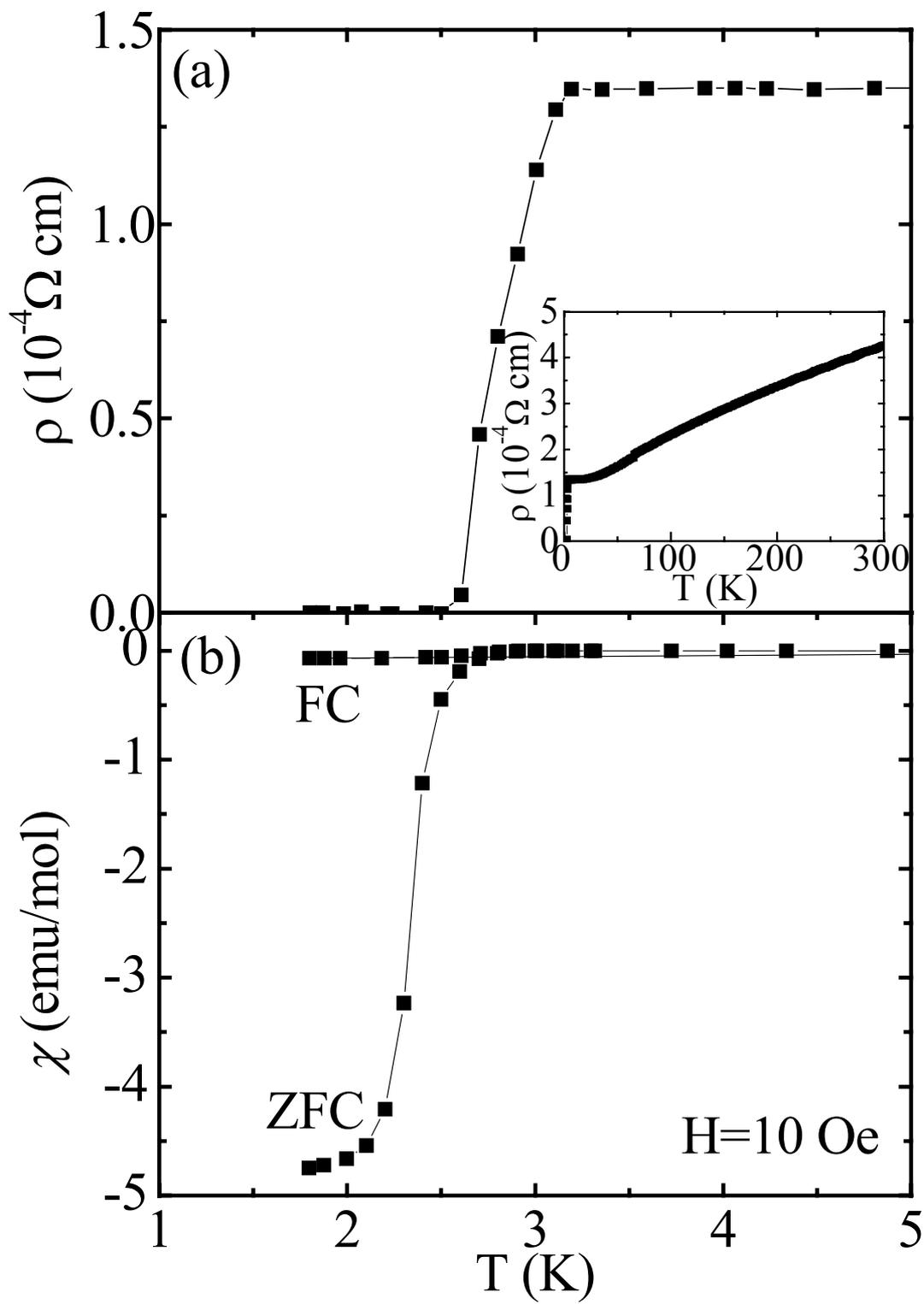

Fig. 2

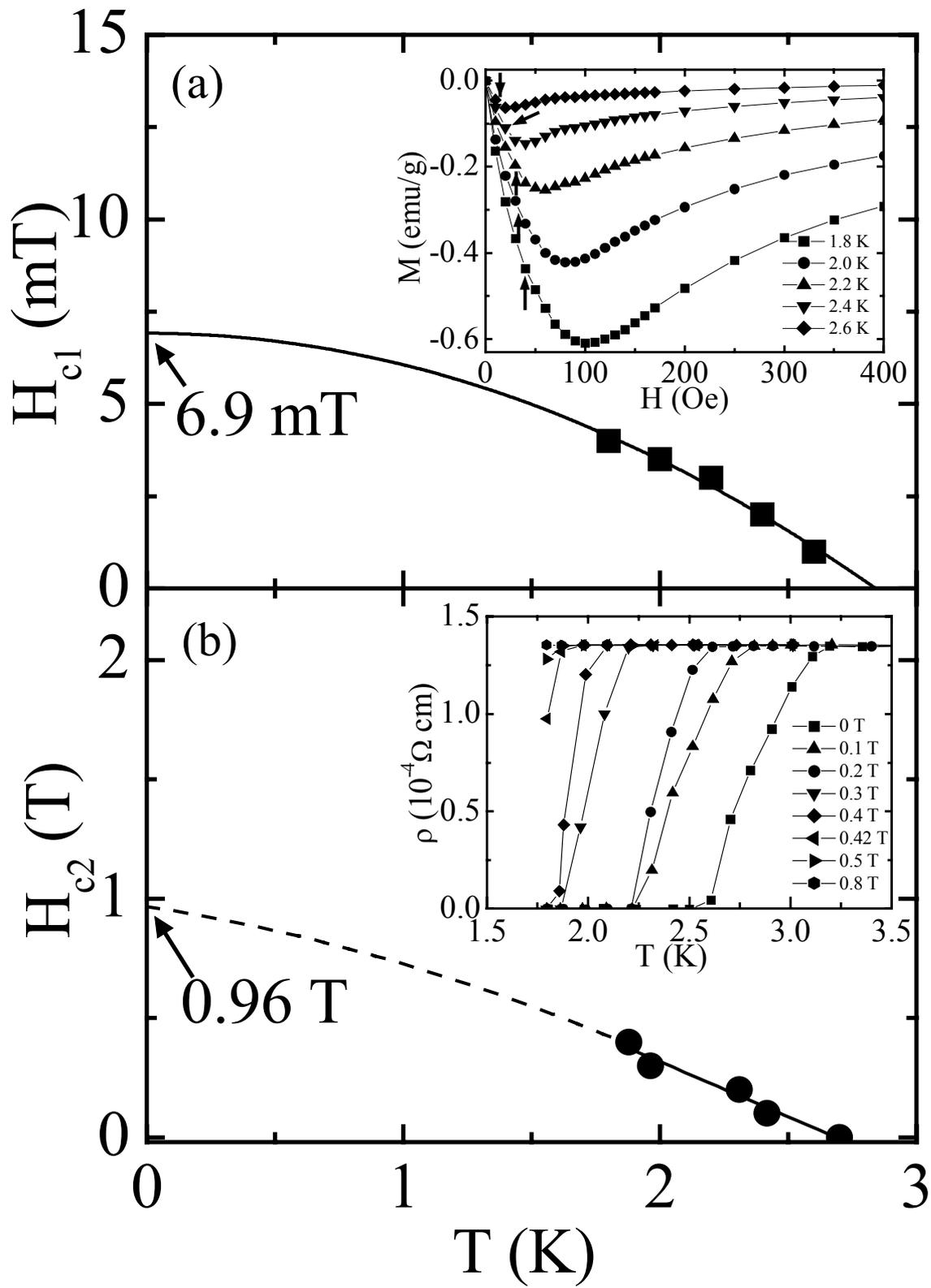

Fig. 3

Table I. Superconducting parameters of ZnN$_y$Ni$_3$, CdCNi$_3$ and MgCNi$_3$.

| Parameters | | MgCNi$_3$ | CdCNi$_3$<br>The values of<br>Sample A in ref. 18. | ZnN$_y$Ni$_3$ |
|---|---|---|---|---|
| Lattice parameter | Å | 3.812 | 3.844 | 3.756 |
| $T_C$ | K | 7.6 | 3.2 | 3.0 |
| $H_{C1}(0)$ | mT | 10 | 8.6 | 6.9 |
| $H_{C2}(0)$ | T | 14.4 | 2.2 | 0.96 |
| $\xi(0)$ | Å | 46 | 122 | 185 |
| $\lambda(0)$ | Å | 2480 | 2767 | 3089 |
| $\kappa(0)$ | | 54 | 23 | 17 |